\documentclass[%
print,
 amsmath,amssymb,
 aps,
]{revtex4}

\usepackage{times}
\usepackage{graphicx}
\usepackage{dcolumn}
\usepackage{bm}
\usepackage{dsfont}
\usepackage{mathrsfs}
\usepackage[landscape,papersize={297.1mm,210mm},left=1.9cm,right=1.6cm,top=2.7cm,bottom=2.8cm]{geometry}
\usepackage[                   
            pdfstartview=FitH,
            colorlinks, 
            pdfborder=001,   
            linkcolor=blue,
            anchorcolor=blue,
            citecolor=blue,
            urlcolor=blue
            ]{hyperref}
\usepackage{amsthm}
\makeatletter

\newcommand{\Rmnum}[1]{\expandafter\@slowromancap\romannumeral #1@}

\makeatother

\begin{document}

\title{Controlled teleportation of an unknown qutrit and its control power}

\author{Xiao-Xu Li$^{1}$}
\author{Feng-Li Yan$^{1}$}
\email{flyan@hebtu.edu.cn}
\author{Ting Gao$^{2}$}
\email{gaoting@hebtu.edu.cn}

\affiliation {$^1$ College of Physics, Hebei Key Laboratory of Photophysics Research and Application, Hebei Normal University, Shijiazhuang 050024, China\\
$^2$ School of Mathematical Sciences, Hebei Normal University, Shijiazhuang 050024, China}

\begin{abstract}
We propose a controlled quantum teleportation protocol for an arbitrary unknown qutrit using a nonmaximally entangled channel, with Charlie holding a qubit and Alice and Bob each holding a four-level particle. With the aid of an auxiliary qubit, the receiver can recover the input state probabilistically with unit fidelity. We derive the overall success probability and show that it is independent of the input state. To quantify the controller's authority, we obtain analytical expressions for the optimal average teleportation fidelities with and without the controller's cooperation, keeping the sender's joint measurement fixed and averaging uniformly over pure qutrit input states with all measurement outcomes included. Their difference defines the control power of the protocol. For the shared states considered here, we determine a controller measurement basis that simultaneously maximizes the control power and the teleportation success probability for a given shared state. These results clarify the role of the controller's measurement in controlled teleportation with unequal local dimensions.~\\


\end{abstract}

\maketitle

\section{Introduction}

Quantum entanglement \cite{1,2,3,4} is an essential resource for a wide range of quantum information tasks, including quantum teleportation \cite{5,6,7,8}, quantum key distribution \cite{9,10,11,12}, and quantum dense coding \cite{13,14,15}. Quantum teleportation, one of its most remarkable applications, was introduced by Bennett et al. \cite{5}. In the standard teleportation protocol, Alice and Bob share a maximally entangled Bell state as their quantum channel. Together with classical communication, this shared entanglement enables the perfect reconstruction of an unknown qubit state at Bob's location without physically transmitting the quantum system initially holding the state. Since its introduction, quantum teleportation has been extensively studied both theoretically and experimentally \cite{5,6,7,8,16,17,18,19,20,21}. Over the past few decades, numerous teleportation protocols employing different entangled resources have been developed, extending the original bipartite single-qubit scheme to probabilistic \cite{22,23,24,25,26} and controlled quantum teleportation \cite{27,28,29,30,31,32}, as well as protocols involving multiple parties \cite{33,34} and high-dimensional quantum systems \cite{31,34,35,36,37}.

In 1998, Karlsson and Bourennane \cite{27} investigated quantum teleportation using a three-particle entangled state, in which reconstruction of the input state requires measurement information supplied by a third participant. Their work provided an early foundation for controlled quantum teleportation. Subsequent studies have extended this setting to bidirectional transmission, multihop networks, and high-dimensional systems. Taufiqi et al. \cite{38} proposed a deterministic bidirectional controlled teleportation protocol using a nonmaximally entangled five-qubit channel. Farimani and Mafi \cite{39} developed a controlled teleportation framework for multihop quantum networks and implemented the protocol on IBM quantum hardware. More recently, Liu et al. \cite{40} proposed a probabilistic multi-output controlled teleportation scheme using a partially entangled high-dimensional channel, allowing multiple unknown single-qudit states to be transmitted to different receivers with the controller's cooperation. These developments motivate further investigation of how the shared entangled resource and the controller's measurement affect teleportation performance.

In quantum teleportation, the local Hilbert space dimensions of the channel particles need not match the dimension of the input system \cite{41,42}. Zhan et al. \cite{43} proposed a scheme for teleporting an unknown qubit state through a high-dimensional entangled channel.  Lei et al. \cite{44} developed a  scheme for an arbitrary unknown two-qubit state via arbitrary high-dimensional entangled states and derived the success probability for the nonmaximally entangled case. High-dimensional controlled teleportation has also been demonstrated experimentally: Lv et al. \cite{31} realized three-dimensional controlled quantum teleportation. These studies motivate the investigation of controlled teleportation in systems with different local dimensions, including the transmission of a qutrit. Beyond teleportation fidelity and success probability, the controller's authority \cite{45} is an important consideration in evaluating controlled quantum teleportation. This authority can be quantified by the difference between the average teleportation fidelities with and without the controller's cooperation, a quantity referred to as control power. Li and Ghose \cite{46} investigated control power for perfect controlled teleportation using partially entangled three-qubit channels. Jeong et al. \cite{47} extended the analysis to more general controlled teleportation settings, defined control power through the difference between the assisted and unassisted average teleportation fidelities, and introduced minimal control power. Subsequent studies examined its relations with tripartite entanglement and teleportation fidelity \cite{48}, as well as its upper bound and trade-off with quantum channel capacity for high-dimensional GHZ and GHZ-type channels \cite{32}. Choi et al. \cite{49} studied the controlled teleportation capability of multipartite qudit states and obtained fidelity thresholds associated with different entanglement structures. More recently, Liu and Mo \cite{50} analyzed controlled teleportation fidelity and minimal control power for parameterized three-particle pure-state channels. These studies motivate us to examine how the controller's measurement affects both the control power and the teleportation success probability, and whether the same measurement basis can maximize both quantities for a given shared entangled state.

In this work, we propose a controlled probabilistic quantum teleportation protocol for an arbitrary unknown qutrit using a nonmaximally entangled tripartite $2\otimes4\otimes4$ channel. Previous studies have investigated high-dimensional controlled bidirectional teleportation using five-particle entangled resources \cite{51} and control power using GHZ and GHZ-type channels \cite{32}. We consider a tripartite channel with unequal local dimensions and study both the teleportation success probability and the control power. We derive an analytical expression for the overall teleportation success probability and show that it is independent of the input state, with unit fidelity achieved upon successful teleportation. By comparing the average teleportation fidelities with and without Charlie's cooperation, we obtain an analytical expression for the control power. Furthermore, for the shared entangled states considered in this work, we show that, with Alice's joint measurement fixed, the same choice of Charlie's measurement basis maximizes both the control power and the teleportation success probability for a given shared state.

This paper is organized as follows. In Sec. \ref{Q2}, we present a controlled probabilistic quantum teleportation protocol using a tripartite entangled channel, describe the corresponding measurement and recovery operations, and derive the overall teleportation success probability. In Sec. \ref{Q3}, we quantitatively analyze the control power by calculating the average teleportation fidelities with and without Charlie's cooperation. We further investigate how Charlie's choice of measurement basis affects the control power and teleportation success probability. Finally, the main results are discussed and summarized in Sec. \ref{Q4}. Our work offers a theoretical perspective on the design and performance analysis of controlled quantum teleportation protocols.

\section{Controlled teleportation of an arbitrary qutrit state}\label{Q2}

We consider a controlled quantum teleportation protocol in which Charlie holds a qubit, while Alice and Bob each hold a 4-level particle. The three parties share the following tripartite entangled channel:
\begin{equation}\label{channel}
|\Psi\rangle_{123}
=a|0\rangle_1\sum_{k=0}^{3}A_{k}|kk\rangle_{23}+b|1\rangle_1\sum_{k=0}^{3}B_{k}|kk\rangle_{23}.
\end{equation}
Here, particle 1 is held by Charlie, while particles 2 and 3 are held by Alice and Bob, respectively, and the real coefficients $a$ and $b$ satisfy $a^{2}+b^{2}=1$. For the two computational basis branches of Charlie, there exist corresponding bipartite entangled states shared by Alice and Bob. Accordingly, the Schmidt coefficients $A_{k}$ and $B_{k}$ can be chosen to be non-negative real number, and are assumed to satisfy the normalization conditions $\sum_{k=0}^{3}A_{k}^{2}=1,\sum_{k=0}^{3}B_{k}^{2}=1$. Suppose that the sender Alice wants to teleport an arbitrary unknown qutrit state of information particle $a$,
\begin{equation}\label{Base}
|\psi\rangle_{a}=\alpha|0\rangle_{a}+\beta|1\rangle_{a}+\gamma|2\rangle_{a}
\end{equation}
to the receiver Bob, where Alice and Bob are in the different places, $\alpha, \beta$ and $\gamma$ are complex numbers, and satisfy the normalization condition $|\alpha|^{2}+|\beta|^{2}+|\gamma|^{2}=1$.

Charlie performs a projective measurement on particle 1 in the basis
\begin{equation}
\begin{aligned}
|\chi_{0}\rangle_{1}
&=\cos\frac{\theta}{2}|0\rangle_{1}+{\mathrm e^{\mathrm i\varphi}}\sin\frac{\theta}{2}|1\rangle_{1},\\
|\chi_{1}\rangle_{1}
&=\sin\frac{\theta}{2}|0\rangle_{1}-{\mathrm e^{\mathrm i\varphi}}\cos\frac{\theta}{2}|1\rangle_{1},
\end{aligned}
\end{equation}
where $0\leq\theta\leq\pi$ and $0\leq\varphi\leq2\pi$. After the measurement, Charlie broadcasts his measurement outcome to Alice and Bob via classical communication. After Charlie's measurement, the shared channel is projected onto $|\Phi_{0}\rangle_{23}$ with probability $p_{0}$, and onto $|\Phi_{1}\rangle_{23}$ with probability $p_{1}$. Therefore, the channel can be rewritten by taking
\begin{equation}
\begin{aligned}
|\Psi\rangle_{123}&=\sqrt{p_{0}}|\chi_{0}\rangle_{1}|\Phi_{0}\rangle_{23}+\sqrt{p_{1}}|\chi_{1}\rangle_{1}|\Phi_{1}\rangle_{23}\\
&=\sqrt{p_{0}}|\chi_{0}\rangle_{1}\sum_{k=0}^{3}d_{0k}|kk\rangle_{23}
+\sqrt{p_{1}}|\chi_{1}\rangle_{1}\sum_{k=0}^{3}d_{1k}|kk\rangle_{23},
\end{aligned}
\end{equation}
where
\begin{equation}\nonumber
\begin{aligned}
&d_{0k}=a A_k \cos\frac{\theta}{2} +b B_k{\mathrm e^{-\mathrm i\varphi}}\sin\frac{\theta}{2},\\
&d_{1k}=a A_k \sin\frac{\theta}{2} -b B_k{\mathrm e^{-\mathrm i\varphi}}\cos\frac{\theta}{2},\\
p_{0}=\sum_{k=0}^{3}|d_{0k}|^{2}&=a^{2}\cos^{2}\frac{\theta}{2}+b^{2}\sin^{2}\frac{\theta}{2}
+ab\sin\theta\cos\varphi\sum_{k=0}^{3}A_{k}B_{k},\\
p_{1}=\sum_{k=0}^{3}|d_{1k}|^{2}&=a^{2}\sin^{2}\frac{\theta}{2}+b^{2}\cos^{2}\frac{\theta}{2}
-ab\sin\theta\cos\varphi\sum_{k=0}^{3}A_{k}B_{k},
\end{aligned}
\end{equation}
and
\begin{equation}\label{B1}
\begin{aligned}
|\Phi_{0}\rangle_{23}=\frac{1}{\sqrt{p_{0}}}\sum_{k=0}^{3}d_{0k}|kk\rangle_{23},\\
|\Phi_{1}\rangle_{23}=\frac{1}{\sqrt{p_{1}}}\sum_{k=0}^{3}d_{1k}|kk\rangle_{23}.
\end{aligned}
\end{equation}
For later convenience, Eq. (\ref{B1}) can be written as
\begin{equation}
\begin{aligned}
|\Phi_{i}\rangle_{23}=\sum_{k=0}^{3}q_{ik}|kk\rangle_{23},
\end{aligned}
\end{equation}
where $q_{ik}=\frac{d_{ik}}{\sqrt{p_{i}}} (i=0,1, k=0,1,2,3),$ are the coefficients of the shared channel associated with Charlie's \(i\)-th measurement outcome.

Alice then performs a joint measurement on particles $a$ and 2 in the following basis
\begin{equation}\label{B2}
\begin{aligned}
|\Omega_{rs}\rangle_{a2}=\frac{1}{\sqrt{3}}\sum_{l=0}^{2}\omega^{rl}|l\rangle_{a}|s\oplus_{4}l\rangle_{2},
\end{aligned}
\end{equation}
where $r=0,1,2$, $s=0,1,2,3$, $\omega=\mathrm{e}^{\mathrm{i}\frac{2\pi}{3}}$ is the triple root, and $\oplus_{4}$ denotes addition modulo 4. Thus, the set $\{|\Omega_{rs}\rangle_{a2}\}$ forms a complete orthonormal basis in the $3\otimes4$-dimensional Hilbert space of particles $a$ and 2, satisfying $_{a2}\langle\Omega_{rs}|\Omega_{r^{'}s^{'}}\rangle_{a2}=\delta_{rr^{'}}\delta_{ss^{'}}$.
It should be noted that this measurement basis is independent of Charlie's measurement outcome and will be used for both conditional channels.

We first consider the case in which Charlie obtains the measurement outcome $|\chi_{0}\rangle_{1}$. In this case, the channel shared by Alice and Bob is projected onto
\begin{equation}
\begin{aligned}
|\Phi_{0}\rangle_{23}=\sum_{k=0}^{3}q_{0k}|kk\rangle_{23}.
\end{aligned}
\end{equation}
Then, the joint state of Alice's unknown particle $a$ and the particles 2 and 3 is
\begin{align}
\nonumber
|\psi\rangle_{a}|\Phi_{0}\rangle_{23}&=(\alpha|0\rangle_{a}+\beta|1\rangle_{a}+\gamma|2\rangle_{a})\otimes\sum_{k=0}^{3}q_{0k}|kk\rangle_{23})\\
&=[\alpha(q_{00}|00\rangle_{a2}|0\rangle_{3}+q_{01}|01\rangle_{a2}|1\rangle_{3}+q_{02}|02\rangle_{a2}|2\rangle_{3}+q_{03}|03\rangle_{a2}|3\rangle_{3})\\
\nonumber
&~~~~+\beta(q_{00}|10\rangle_{a2}|0\rangle_{3}+q_{01}|11\rangle_{a2}|1\rangle_{3}+q_{02}|12\rangle_{a2}|2\rangle_{3}+q_{03}|13\rangle_{a2}|3\rangle_{3})\\
\nonumber
&~~~~+\gamma(q_{00}|20\rangle_{a2}|0\rangle_{3}+q_{01}|21\rangle_{a2}|1\rangle_{3}+q_{02}|22\rangle_{a2}|2\rangle_{3}+q_{03}|23\rangle_{a2}|3\rangle_{3})].
\end{align}
After measuring particle $a$ and particle 2 in Alice's hand using the basis (\ref{B2}), the unnormalized collapsed states of particle 3 corresponding to Alice's joint measurement basis states are $|\phi_{j}\rangle_{3}=_{a2}\langle\Omega_{rs}(|\psi\rangle_{a}|\Phi_{0}\rangle_{23})$, which are
\begin{equation}\label{B3}
\begin{aligned}
|\zeta_{0rs}\rangle_{3}=\frac{1}{\sqrt{3}}(\alpha q_{0,s}|s\rangle_{3}+\beta\omega^{-r}q_{0,s\oplus_{4}1}|s\oplus_{4}1\rangle_{3}
+\gamma\omega^{-2r}q_{0,s\oplus_{4}2}|s\oplus_{4}2\rangle_{3}),
\end{aligned}
\end{equation}
where $r=0,1,2$, $s=0,1,2,3$. For clarity, the twelve possible states of Bob's particle corresponding to Alice's measurement outcomes are explicitly listed below. When $s=0$, one obtains
\begin{equation}\label{B4}
\begin{aligned}
&|\zeta_{000}\rangle_{3}=\frac{1}{\sqrt{3}}(\alpha q_{00}|0\rangle_{3}+\beta q_{01}|1\rangle_{3}+\gamma q_{02}|2\rangle_{3},\\
&|\zeta_{010}\rangle_{3}=\frac{1}{\sqrt{3}}(\alpha q_{00}|0\rangle_{3}+\beta\omega^{2} q_{01}|1\rangle_{3}+\gamma\omega q_{02}|2\rangle_{3},\\
&|\zeta_{020}\rangle_{3}=\frac{1}{\sqrt{3}}(\alpha q_{00}|0\rangle_{3}+\beta\omega q_{01}|1\rangle_{3}+\gamma\omega^{2}q_{02}|2\rangle_{3}.
\end{aligned}
\end{equation}
When $s=1$, the corresponding states are
\begin{equation}\label{B5}
\begin{aligned}
&|\zeta_{001}\rangle_{3}=\frac{1}{\sqrt{3}}(\alpha q_{01}|1\rangle_{3}+\beta q_{02}|2\rangle_{3}+\gamma q_{03}|3\rangle_{3},\\
&|\zeta_{011}\rangle_{3}=\frac{1}{\sqrt{3}}(\alpha q_{01}|1\rangle_{3}+\beta\omega^{2} q_{02}|2\rangle_{3}+\gamma\omega q_{03}|3\rangle_{3},\\
&|\zeta_{021}\rangle_{3}=\frac{1}{\sqrt{3}}(\alpha q_{01}|1\rangle_{3}+\beta\omega q_{02}|2\rangle_{3}+\gamma\omega^{2}q_{03}|3\rangle_{3}.
\end{aligned}
\end{equation}
When $s=2$, one has
\begin{equation}\label{B6}
\begin{aligned}
&|\zeta_{002}\rangle_{3}=\frac{1}{\sqrt{3}}(\alpha q_{02}|2\rangle_{3}+\beta q_{03}|3\rangle_{3}+\gamma q_{00}|0\rangle_{3},\\
&|\zeta_{012}\rangle_{3}=\frac{1}{\sqrt{3}}(\alpha q_{02}|2\rangle_{3}+\beta\omega^{2} q_{03}|3\rangle_{3}+\gamma\omega q_{00}|0\rangle_{3},\\
&|\zeta_{022}\rangle_{3}=\frac{1}{\sqrt{3}}(\alpha q_{02}|2\rangle_{3}+\beta\omega q_{03}|3\rangle_{3}+\gamma\omega^{2}q_{00}|0\rangle_{3}.
\end{aligned}
\end{equation}
Finally, when $s=3$, the corresponding states are given by
\begin{equation}\label{B7}
\begin{aligned}
&|\zeta_{003}\rangle_{3}=\frac{1}{\sqrt{3}}(\alpha q_{03}|3\rangle_{3}+\beta q_{00}|0\rangle_{3}+\gamma q_{01}|1\rangle_{3},\\
&|\zeta_{013}\rangle_{3}=\frac{1}{\sqrt{3}}(\alpha q_{03}|3\rangle_{3}+\beta\omega^{2} q_{00}|0\rangle_{3}+\gamma\omega q_{01}|1\rangle_{3},\\
&|\zeta_{023}\rangle_{3}=\frac{1}{\sqrt{3}}(\alpha q_{03}|3\rangle_{3}+\beta\omega q_{00}|0\rangle_{3}+\gamma\omega^{2}q_{01}|1\rangle_{3}.
\end{aligned}
\end{equation}

According to Alice's measurement outcome $(r,s)$, Bob first applies the following cyclic shift operation to particle 3 :
\begin{equation}\label{B8}
\begin{aligned}
U_{s}=\sum_{l=0}^{3}|l\rangle_{3}\langle s\oplus_{4}l|.
\end{aligned}
\end{equation}
Bob then applies the phase operation :
\begin{equation}\label{B9}
\begin{aligned}
V_{r}=|0\rangle_{3}\langle0|+\omega^{r}|1\rangle_{3}\langle1|+\omega^{2r}|2\rangle_{3}\langle2|+|3\rangle_{3}\langle3|.
\end{aligned}
\end{equation}
This operation removes the phase factors introduced by Alice's measurement basis. Consequently, Bob obtains
\begin{equation}\label{B10}
\begin{aligned}
|\eta_{0rs}\rangle=V_{r}U_{s}|\zeta_{0rs}\rangle_{3}=\frac{1}{\sqrt{3}}(\alpha q_{0,s}|0\rangle_{3}+\beta q_{0,s\oplus_{4}1}|1\rangle_{3}
+\gamma q_{0,s\oplus_{4}2}|2\rangle_{3}).
\end{aligned}
\end{equation}

Depending on Alice's measurement outcome, Bob obtains the corresponding state of particle 3 and subsequently applies the appropriate local operations to recover the unknown state. Probabilistically from the state of particle 3, Bob introduces an auxiliary qubit $b$, initially prepared in the state $|0\rangle_{b}$, and performs a joint unitary operation on particle 3 and the auxiliary qubit. After the auxiliary qubit $b$ is initialized in the state $|0\rangle_{b}$, their joint state can be expressed as
\begin{equation}\label{B11}
\begin{aligned}
|\Xi\rangle_{3b}&=|\eta_{0rs}\rangle_{3}|0\rangle_{b}\\
&=\frac{(\alpha q_{0,s}|0\rangle_{3}|0\rangle_{b}+\beta q_{0,s\oplus_{4}1}|1\rangle_{3}|0\rangle_{b}
+\gamma q_{0,s\oplus_{4}2}|2\rangle_{3}|0\rangle_{b})}{\sqrt{|\alpha|^{2}|q_{0,s}|^{2}+|\beta|^{2}|q_{0,s\oplus_{4}1}|^{2}+|\gamma|^{2}|q_{0,s\oplus_{4}2}|^{2}}}.
\end{aligned}
\end{equation}
The three relevant channel coefficients can be expressed in polar form as $q_{0,s\oplus_{4}j}=|q_{0,s\oplus_{4}1}|\mathrm{e}^{\mathrm{i}\delta_{j}}$ $(j=0,1,2)$, where $\delta_{j}=\mathrm{arg}(q_{0,s\oplus_{4}j})$ and $0\leq\delta_{j}<2\pi$. The minimum magnitude among these coefficients is denoted by $q_{\mathrm
{min}}^{(0,s)}=\mathrm{min}\{|q_{0,s}|, |q_{0,s\oplus_{4}1}|, |q_{0,s\oplus_{4}2}|\}$. To ensure the successful recovery of an arbitrary unknown qutrit state in this branch, all three coefficients $q_{0,s}, q_{0,s\oplus_{4}1}$, and $q_{0,s\oplus_{4}2}$ are required to be nonzero. In order the basis $\{|00\rangle_{3b}, |01\rangle_{3b}, |10\rangle_{3b}, |11\rangle_{3b}, |20\rangle_{3b}, |21\rangle_{3b}\}$, the joint unitary transformation $U_{3b}^{0,s}$ can be chosen as
\begin{equation}\label{B12}
\begin{aligned}
U_{3b}^{0,s}=R_{0}^{(0,s)}\oplus R_{1}^{(0,s)}\oplus R_{2}^{(0,s)}\oplus I_{2}.
\end{aligned}
\end{equation}
For $j=0,1,2$, the $2\times2$ block $R_{j}^{(0,s)}$ acts on the subspace spanned by $ \{|j0\rangle_{3b},|j1\rangle_{3b}\}$, and is given by
\begin{equation}\label{B13}\nonumber
\begin{aligned}
R_{j}^{(0,s)} =\left(\begin{array}{cc}
        \mathrm{e}^{-\mathrm{i}\delta_{j}}\frac{q_{\mathrm{min}}^{(0,s)}}{|q_{0,s\oplus_{4}j}|}
& -\sqrt{1-\frac{(q_{\mathrm{min}}^{(0,s)})^2}{|q_{0,s\oplus_{4}j}|^{2}}}\\
        \sqrt{1-\frac{(q_{\mathrm{min}}^{(0,s)})^2}{|q_{0,s\oplus_{4}j}|^{2}}}
 & \mathrm{e}^{\mathrm{i}\delta_{j}}\frac{q_{\mathrm{min}}^{(0,s)}}{|q_{0,s\oplus_{4}j}|}
\end{array}
\right),
\end{aligned}
\end{equation}
where the identity matrix $I_{2}$ acts on the subspace spanned by the two basis states $\{|30\rangle_{3b},|31\rangle_{3b}\}$. Applying $U_{3b}^{0,s}$ to the joint state in Eq. (\ref{B11}) gives
\begin{equation}\label{B14}
\begin{aligned}
|\Gamma\rangle&=U_{3b}^{0,s}\frac{(\alpha q_{0,s}|0\rangle_{3}|0\rangle_{b}+\beta q_{0,s\oplus_{4}1}|1\rangle_{3}|0\rangle_{b}
+\gamma q_{0,s\oplus_{4}2}|2\rangle_{3}|0\rangle_{b})}{\sqrt{|\alpha|^{2}|q_{0,s}|^{2}+|\beta|^{2}|q_{0,s\oplus_{4}1}|^{2}+|\gamma|^{2}|q_{0,s\oplus_{4}2}|^{2}}}\\
&=\frac{q_{\mathrm{min}}^{(0,s)}}{\sqrt{|\alpha|^{2}|q_{0,s}|^{2}+|\beta|^{2}|q_{0,s\oplus_{4}1}|^{2}+|\gamma|^{2}|q_{0,s\oplus_{4}2}|^{2}}}(\alpha|0\rangle_{3}+\beta|1\rangle_{3}+\gamma|2\rangle_{3})|0\rangle_{b}\\
&~~~~~+\frac{(\alpha q_{0,s}\sqrt{1-\frac{(q_{\mathrm{min}}^{(0,s)})^{2}}{|q_{0,s}|^{2}}}|0\rangle_{3}+\beta q_{0,s\oplus_{4}1}\sqrt{1-\frac{(q_{\mathrm{min}}^{(0,s)})^{2}}{|q_{0,s\oplus_{4}1}|^{2}}}|1\rangle_{3}+\gamma q_{0,s\oplus_{4}2}\sqrt{1-\frac{(q_{\mathrm{min}}^{(0,s)})^{2}}{|q_{0,s\oplus_{4}2}|^{2}}}|2\rangle_{3})|1\rangle_{b}}{\sqrt{|\alpha|^{2}|q_{0,s}|^{2}+|\beta|^{2}|q_{0,s\oplus_{4}1}|^{2}+|\gamma|^{2}|q_{0,s\oplus_{4}2}|^{2}}}.
\end{aligned}
\end{equation}
Subsequently, one performs a von Neumann measurement on the introduced auxiliary particle $b$. If the measurement outcome of auxiliary particle $b$ is $|0\rangle_{b}$, particle 3 collapses to the original unknown state (\ref{Base}), the teleportation is successful; if the measurement result is $|1\rangle_{b}$, the original unknown state cannot be recovered from the resulting state of particle 3, and the teleportation attempt is regarded as unsuccessful.

Conditioned on Charlie obtaining outcome $|0\rangle_{b}$ and Alice obtaining outcome $(r,s)$, the success probability is
\begin{equation}\label{B15}
\begin{aligned}
P_{0rs}=\frac{(q_{\mathrm{min}}^{(0,s)})^{2}}{|\alpha|^{2}|q_{0,s}|^{2}+|\beta|^{2}|q_{0,s\oplus_{4}1}|^{2}+|\gamma|^{2}|q_{0,s\oplus_{4}2}|^{2}}
\end{aligned}.
\end{equation}
Next we will investigate the total success probability of the teleportation. For Charlie's measurement outcome, the total success probability associated with the channel $|\Phi_{0}\rangle_{23}$ is given by
\begin{equation}\label{B16}
\begin{aligned}
P^{(1)}=\sum_{s=0}^{3}(q_{\mathrm{min}}^{(0,s)})^{2}
\end{aligned}.
\end{equation}

When Charlie obtains the measurement outcome $|\chi_{1}\rangle_{1}$, the entangled state shared by Alice and Bob is $|\Phi_{1}\rangle_{23}\sum_{k=0}^{3}q_{1k}|kk\rangle_{23}$. The subsequent teleportation procedure is analogous to that for $|\Phi_{1}\rangle_{23}$. By replacing  $q_{0k}$ with $q_{1k}$ in the corresponding operations, Bob can recover the original unknown qutrit state with unit fidelity when the measurement outcome of the auxiliary qubit is $|0\rangle_{b}$. Let $q_{\mathrm{min}}^{(1,s)}=\mathrm{min}\{|q_{1,s}|, |q_{1,s\oplus_4 1}|, |q_{1,s\oplus_4 2}|\}$. The success probability corresponding to Charlie's measurement outcome $|\chi_{1}\rangle_{1}$, summed over all possible measurement outcomes of Alice, is given by
\begin{equation}\label{B17}
\begin{aligned}
P^{(2)}=\sum_{s=0}^{3}(q_{\mathrm{min}}^{(1,s)})^{2}
\end{aligned}.
\end{equation}

Taking into account Charlie's two possible measurement outcomes, the total success probability of the teleportation protocol can be expressed as
\begin{equation}\label{B18}
\begin{aligned}
P_{\mathrm{total}}=\sum_{i=0}^{1}\sum_{s=0}^{3}\mathrm{min}\{|d_{i,s}|^{2}, |d_{i,s\oplus_{4}1}|^{2}, |d_{i,s\oplus_{4}2}|^{2}\}
\end{aligned}.
\end{equation}
This result shows that the total success probability depends on the coefficients of the shared entangled state and Charlie's measurement parameters, but is independent of the unknown input qutrit state. For each successful outcome, Bob recovers the original unknown state with unit fidelity.

\section{Control power of the protocol}\label{Q3}

We examine Charlie's role by comparing the average teleportation fidelities achieved with and without his assistance. Unlike the probabilistic recovery in Sec. \ref{Q2}, where the input state is recovered exactly upon a successful auxiliary qubit measurement, the average fidelity considered here includes all outcomes. We then investigate how Charlie's measurement basis affects both the control power and the success probability of the original protocol.

\subsection{Teleportation fidelity and control power}

We consider the shared state in Eq. (\ref{channel}) and Alice's joint measurement in Eq. (\ref{B2}). When Charlie communicates his measurement outcome, Bob can use both Charlie's and Alice's outcomes to choose his recovery operation. Without Charlie's assistance, Bob's recovery depends only on Alice's outcome. Alice's joint measurement remains unchanged in both cases, and Bob's recovery operations are deterministic.

To calculate the average teleportation fidelity, we assume that the pure qutrit input states are uniformly distributed \cite{52,53,54}. The average fidelity is then given by
\begin{equation}\label{B19}
\begin{aligned}
\bar{F}=\int d\psi F(\psi),
\end{aligned}
\end{equation}
where, $d\psi$ is the normalized unitarily invariant measure on the space of pure qutrit states, with $\int d\psi=1$. Following the definition of control power in Ref. \cite{47}, we quantify Charlie's contribution by the difference between the average teleportation fidelities with and without his assistance:
\begin{equation}\label{B20}
\begin{aligned}
\xi=\bar{F}_{C}-\bar{F}_{NC}.
\end{aligned}
\end{equation}
Here, $\bar{F}_{C}$ and $\bar{F}_{NC}$ denote the maximum average teleportation fidelities with and without Charlie's assistance, for the prescribed Alice measurement and the given choice of Charlie's measurement basis. Bob chooses the corresponding quantum operations based on the available measurement outcomes to maximize the average fidelity, with all outcomes included in the average.

We first consider the case in which Charlie does not communicate his measurement outcome. The state shared by Alice and Bob is then obtained by tracing out particle 1:
\begin{equation}\label{B21}
\begin{aligned}
\rho_{23}^{NC}&=\mathrm{Tr}_{1}|\Psi\rangle_{123}\langle\Psi|\\
&=\sum_{k,l=0}^{3}G_{kl}|kk\rangle_{23}\langle ll|,
\end{aligned}
\end{equation}
where $G_{kl}=a^{2}A_{k}A_{l}+b^{2}B_{k}B_{l}$. Since $A_{k}$ and $B_{k}$ are real and nonnegative, $G_{kl}\geq0$. The shared state remains normalized, with $\sum_{k=0}^{3}G_{kk}=1$. When Charlie's measurement outcome is not communicated, averaging over all his possible outcomes gives the same reduced state $\rho_{23}^{NC}$, irrespective of his measurement basis. For convenience, we write the coefficients of the input state by $c_{0}=\alpha$, $c_{1}=\beta$, and $c_{2}=\gamma$. Based on Alice's measurement outcome, Bob successively applies $U_{s}$ and $V_{r}$, given in Eqs. (\ref{B8}) and (\ref{B9}). The resulting  state is
\begin{equation}\label{B22}
\begin{aligned}
\rho^{NC}=\sum_{j,m=0}^{2}\sum_{s=0}^{3}c_{j}c_{m}^{\ast}G_{s\oplus_{4}j,s\oplus_{4}m}|j\rangle_{3}\langle m|.
\end{aligned}
\end{equation}
For a given measurement outcome, the teleportation fidelity is expressed as
\begin{equation}\label{B22-1}
\begin{aligned}
F&=\langle\psi|\rho^{NC}|\psi\rangle=\sum_{j=0}^{2}c_{j}^{\ast}\langle j|\rho^{NC}|\sum_{m=0}^{2}c_{m}|m\rangle\\
&=\sum_{j,m=0}^{2}|c_{j}|^{2}|c_{m}|^{2}\sum_{s=0}^{3}G_{s\oplus_{4}j,s\oplus_{4}m}.
\end{aligned}
\end{equation}
The probability that Alice obtains the measurement outcome associated with $|\Omega_{rs}\rangle_{a2}$, which is also the probability of obtaining the corresponding state at Bob's side, is
\begin{equation}\label{B23}
\begin{aligned}
P^{NC}=\frac{1}{3}\sum_{j=0}^{2}|c_{j}|^{2}G_{s\oplus_{4}j,s\oplus_{4}j}.
\end{aligned}
\end{equation}

Averaging uniformly over the input state $|\psi\rangle_{a}$, we obtain
\begin{equation}\label{B24}
\begin{aligned}
\bar{F}_{NC}&=\int d\psi\sum_{s=0}^{3}\sum_{j,m=0}^{2}|c_{j}|^{2}|c_{m}|^{2}G_{s\oplus_{4}j,s\oplus_{4}m}\\
&=\frac{1}{3}+\frac{a^{2}(\sum_{k=0}^{3}A_{k})^{2}+b^{2}(\sum_{k=0}^{3}B_{k})^{2}}{6}.
\end{aligned}
\end{equation}

Next, we consider the case in which Charlie communicates his measurement outcome. Averaging uniformly over the input state $|\psi\rangle_{a}$, we obtain
\begin{equation}\label{B25}
\begin{aligned}
\bar{F}_{C}&=\int d\psi\sum_{i=0}^{1}p_{i}\sum_{s=0}^{3}(\sum_{j=0}^{2}|c_{j}|^{2}|q_{i,s\oplus_{4}j}|)^{2}\\
&=\frac{1}{3}+\frac{\sum_{i=0}^{1}p_{i}(\sum_{k=0}^{3}|q_{ik}|)^{2}}{6},
\end{aligned}
\end{equation}
where $p_{i}$ is the probability that Charlie obtains measurement outcome $i$, with $\sum_{i=0}^{1} p_{i}=1$ and $\sum_{k=0}^{3}|q_{ik}|^{2}=1$ for each outcome with $p_{i}>0$. Outcomes with $p_{i}=0$ don't contribute to the average and omitted from the calculation.

Substituting Eqs. (\ref{B24}) and (\ref{B25}) into Eq. (\ref{B20}), we obtain the control power
\begin{equation}\label{B26}
\begin{aligned}
\xi=\frac{1}{6} [\sum_{i=0}^{1}p_{i}(\sum_{k=0}^{3}|q_{ik}|)^{2}-a^{2}(\sum_{k=0}^{3}A_{k})^{2}-b^{2}(\sum_{k=0}^{3}B_{k})^{2}].
\end{aligned}
\end{equation}
Using the normalization conditions, Eq. (\ref{B26}) above can be rewritten as
\begin{equation}\label{B27}
\begin{aligned}
\xi=\frac{1}{3}\sum_{k>l=0}^{3} [\sum_{i=0}^{1}p_{i}(|q_{ik}||q_{il}|)-G_{kl}].
\end{aligned}
\end{equation}
When Charlie does not communicate his measurement outcome to Alice and Bob, their shared channel is $\rho_{23}^{NC}=\sum_{i=0}^{1}p_{i}|\Phi_{i}\rangle_{23}\langle\Phi_{i}|$, which gives $\sum_{i=0}^{1}p_{i}q_{ik}q_{il}^{\ast}=G_{kl}$. Since $G_{kl}\geq0$, the triangle inequality gives $G_{kl}=|\sum_{i=0}^{1}p_{i}q_{ik}q_{il}^{\ast}|\leq\sum_{i=0}^{1}p_{i}|q_{ik}||q_{il}|$. Therefore, $\xi\geq0$.

\subsection{Choice of Charlie's measurement basis and teleportation success probability}

We next examine how Charlie's measurement basis affects the control power and the success probability of the original protocol. For a given shared state in Eq. (\ref{channel}),  $\bar{F}_{NC}$ is independent of Charlie's measurement basis. Maximizing $\xi$ over this basis is therefore equivalent to maximizing $\bar{F}_{C}$. We define
\begin{equation}\label{B28}
\begin{aligned}
t_{k}=G_{kk}=a^{2}A_{k}^{2}+b^{2}B_{k}^{2},
\end{aligned}
\end{equation}
where $\sum_{k=0}^{3}t_{k}=1$, and $\sum_{i=0}^{1}p_{i}|q_{ik}|^{2}=t_{k}$. Using the Cauchy-Schwarz inequality together with Eq. (\ref{B28}), Eq. (\ref{B27}) yields
\begin{equation}\label{B29}
\begin{aligned}
\xi&\leq\frac{1}{3}\sum_{k>l=0}^{3}(\sqrt{t_{k}t_{l}}-G_{kl}).
\end{aligned}
\end{equation}

When $\theta=\varphi=\frac{\pi}{2}$, the upper bound in Eq. (\ref{B29}) is attained. Charlie's measurement basis then takes the form
\begin{equation}\label{B30}
\begin{aligned}
|\chi_{0}\rangle_{1}
&=\frac{|0\rangle_{1}+\mathrm{i}|1\rangle_{1}}{\sqrt{2}},\\
|\chi_{1}\rangle_{1}
&=\frac{|0\rangle_{1}-\mathrm{i}|1\rangle_{1}}{\sqrt{2}}.
\end{aligned}
\end{equation}
For this choice, $q_{0k}=aA_{k}-\mathrm{i}bB_{k}$, and $q_{1k}=aA_{k}+\mathrm{i}bB_{k}$. Since $a, b, A_{k}$ and $B_{k}$ are real, we have $|q_{0k}|^{2}=|q_{1k}|^{2}=t_{k}$. Together with $p_{0}=p_{1}=\frac{1}{2}$, this shows that the chosen measurement basis attains the above upper bound. Therefore, the maximum control power over Charlie's measurement basis is
\begin{equation}\label{B31}
\begin{aligned}
\xi&=\frac{1}{3}\sum_{k>l=0}^{3}(\sqrt{t_{k}t_{l}}-G_{kl}).
\end{aligned}
\end{equation}

We now consider the success probability of the original controlled quantum teleportation protocol, which satisfies
\begin{equation}\label{B32}
\begin{aligned}
P=\sum_{i=0}^{1}p_{i}\sum_{s=0}^{3}\mathrm{min}\{|q_{i,s\oplus_{4}j}|^{2}\}\leq\sum_{s=0}^{3}\mathrm{min}\{t_{s\oplus_{4}j}\},
\end{aligned}
\end{equation}
where $j=0, 1, 2$. The measurement basis chosen above attains this bound, giving
\begin{equation}\label{B33}
\begin{aligned}
P_{\mathrm{max}}=\sum_{s=0}^{3}\mathrm{min}\{t_{s\oplus_{4}j}\}.
\end{aligned}
\end{equation}
Let $t_{1}\leq t_{2}\leq t_{3}\leq t_{4}$ denote the four values of $t_{k}$ in ascending order. Eq. (\ref{B33}) then gives
\begin{equation}\label{B34}
\begin{aligned}
P_{\mathrm{max}}=3t_{1}+t_{2}.
\end{aligned}
\end{equation}
Thus, for a given shared channel state, the same measurement basis simultaneously maximizes the control power and the success probability. When Charlie chooses the measurement basis corresponding to $\theta=\varphi=\frac{\pi}{2}$, both the control power and the success probability of the original protocol attain their respective maximum values.

\section{Discussion and summary}\label{Q4}

In summary, we have presented a controlled teleportation protocol for an arbitrary unknown qutrit state using a tripartite $2\otimes4\otimes4$ entangled channel. Charlie holds a qubit, while Alice and Bob each hold a four-level particle. Following Charlie's measurement and Alice's joint measurement, Bob recovers the input state probabilistically with the aid of an auxiliary qubit. The input state is recovered with unit fidelity whenever the auxiliary qubit measurement indicates success. The overall success probability depends on the shared state coefficients and Charlie's measurement basis, but is independent of the input state.

Previous studies have examined the relations between control power, teleportation fidelity, and multipartite entanglement \cite{47,48}, as well as the upper bound on control power and its trade-off with quantum channel capacity for high-dimensional GHZ and GHZ-type channels \cite{32}. Complementing these studies, our analysis considers both the control power and the probability of exact recovery in the proposed protocol. For Alice's prescribed joint measurement, we have derived the optimal average teleportation fidelities with and without Charlie's cooperation, averaged over uniformly distributed pure qutrit input states with all measurement outcomes included. Their difference gives the control power of the protocol. The noncooperative fidelity is independent of Charlie's measurement basis, whereas the cooperative fidelity can be improved by an appropriate choice of that basis. For the same shared entangled state, we study both the increase in average teleportation fidelity due to Charlie's cooperation and the success probability of the original protocol.

For a given shared channel state considered in this work, we have shown that the measurement basis corresponding to $\theta=\varphi=\frac{\pi}{2}$ simultaneously maximizes the control power and the teleportation success probability. The two measurement outcomes occur with equal probability, and the corresponding conditional channels satisfy $|q_{0k}|^2=|q_{1k}|^2=t_{k}$. If these four values satisfy $t_{1}\leq t_{2}\leq t_{3}\leq t_{4}$, the maximum success probability is $P_{\max}=3t_{1}+t_{2}$. Thus, the two smallest values determine the maximum success probability, while the control power also depends on the off-diagonal coefficients $G_{kl}$ of the reduced shared state. These results clarify the role of Charlie's measurement in determining the control power and success probability of qutrit teleportation through the $2\otimes4\otimes4$ channel.

The present work contributes to the study of controlled quantum teleportation with unequal local dimensions by examining both the teleportation success probability and the control power. This approach may help guide the choice of entangled resources and measurements when both aspects are relevant to protocol design. Extending the analysis to more general shared channel states and higher-dimensional input states would allow us to determine whether, and under what conditions, a single choice of the controller's measurement basis can still maximize both the control power and the teleportation success probability.\\

\begin{acknowledgements}
This work was supported by
the National Natural Science Foundation of China under Grant No. 62271189.\\
\end{acknowledgements}

\noindent{\bf Conflicts of Interest}\\

The authors declare no conflicts of interest.\\

\noindent{\bf Data Availability Statement}\\

Data sharing is not applicable to this article as no new data were created or analyzed in this study.\\

\end{document}